\documentclass[twocolumn,10pt,aps,superscriptaddress,prl]{revtex4}
\usepackage{epsfig}
\usepackage{amsmath}%
\usepackage{amsfonts}%
\usepackage{amssymb}%
\usepackage{graphicx}
\begin{document}
\bibstyle{plain}
\title{Multisetting Bell inequalities for $N$ spin-1 systems avoiding KS contradiction}
\author{Arijit Dutta}\affiliation{Institute of Theoretical Physics and Astrophysics, University of Gda\'nsk, PL-80-952 Gda\'nsk, Poland}
\author{Marcin Wie\'sniak}\affiliation{Institute of Theoretical Physics and Astrophysics, University of Gda\'nsk, PL-80-952 Gda\'nsk, Poland}
\author{Marek \.Zukowski}\affiliation{Institute of Theoretical Physics and Astrophysics, University of Gda\'nsk, PL-80-952 Gda\'nsk, Poland}
\begin{abstract}
Bell's theorem for systems more complicated than two qubits faces a hidden, as yet undiscussed, problem. One of the methods to derive Bell's inequalities is to assume existence of joint probability distribution for measurement results for all settings in the given experiment. However for spin-1 systems, one faces the problem that eigenvalues of observables do not allow a consistent algebra if one allows all possible settings on each side (Bell 1966 contradiction), or some specific sets (leading to a Kochen-Specker 1967 contradiction). We show here that by choosing special set of  settings which never lead to inconsistent algebra of eigenvalues, one can still derive multisetting  Bell inequalities, and that they are robustly violated. Violation factors increase with the number of subsystems. The inequalities involve only spin observables, we do not allow all possible qutrit observables, still the violations are strong.
\end{abstract}

\maketitle
{\em Introduction}: Intrinsic randomness of Quantum Mechanics has always been difficult to accept. It was challenged by Einstein, Podolsky, and Rosen \cite{epr}. Many years later Bell \cite{bell}  formulated  an inequality invalidating this challenge. Violations of Bell-type inequalities in an experiment or theory exclude local and  realistic completions of quantum mechanics. There have been many experimental attempts to violate a Bell inequality, for a recent review see \cite{PAN}, often related to some specific challenges. 
For example, in Refs. \cite{aspect,wineland,weihs} the challenge  was to close various loopholes allowing a Local Realistic description. In the meantime, it turned out that Bell's inequalities are useful tools to classify non-classicality of quantum correlations, which could be applied in various quantum information protocols breaching the limits of classical communication and classical computation. Thus, Bell inequalities have now a practical importance.

We have many generalizations of the inequalities. For more than two qubits   Mermin introduced a series of inequalities for experiments with choice between two dichotomic observables, \cite{mermin}. Alternative versions were given in Refs \cite{ardehali,kb}, and finally, the full set in \cite{ww, wz}, and \cite{zb}. More recently,  a further generalization, to more than two settings per observer, was introduced e.g. in Refs. \cite{chen} and \cite{lpzb}.

In the case of higher-dimensional systems early results can be found in Refs \cite{MERMIN}.
Later,  Kaszlikowski {\em et al.} \cite{dago1}  have provided  an evidence that two entangled qutrits violate Bell inequalities stronger than two entangled qubits. This was confirmed in \cite{dago2}.
 A Clauser-Horne-type \cite{CH} inequality was proposed for two qutrits in Ref. \cite{dago0}. Finally, Collins {\em et al.} have derived series of two-setting two-observer Bell inequalities involving measurements of any finite  dimensionality \cite{collins}, and a different form has been derived by Zohren and Gill, \cite{ZOHREN}.
 A numerical study of more settings scenarios two-observer scenarios can be found in Ref. \cite{GRUCA}.

 A three-dimensional Hilbert space is sufficient to conduct a proof of the Kochen-Specker (KS) theorem \cite{ks}, which falsifies non-contextual hidden variables. Recently, a non-contextual realistic description of a single three-dimensional system was falsified in an experiment \cite{radek}. This may make constructing Bell inequalities for qutrits look pointless. Nevertheless, a closer examination of the problem still provides us with a motivation. The KS theorem shows us that there cannot be hidden variables describing results of {\em all} measurements, as some subsets of measurement cannot be modeled so. Such sets, often called KS sets can be now found algorithmically \cite{PAVICIC}. However, there are {\em certain} other subsets, for which these models are not invalidated by KS-type reasonings. 

Here we introduce multisetting Bell-type inequalities for collections of spin-1 systems, which we shall call cone Bell inequalities. We shall consider only observables  related with spin measurements, this is a completely different approach than in \cite{dago1,dago2,dago0,collins}, where observables with no classical analogues were considered (the specific observables involved multiport beamsplitters). This is the reason why we call our subsystems spin-1 systems. The terminology stresses that we do not allow all possible qutrit observables. The word ``cone" used to describe the inequalities reflects the fact that the Stern-Gerlach-type measurements entering the inequalities have the property that the vectors defining the measured component of the spin can come only from a specific cone (described in more detail further on). The cone sets are such that every projective measurement appears in a single (local) context, thus they cannot form KS sets leading to a contradiction, for whatever the number of settings. One can fully separate the two no-go statements against hidden variables. And perhaps even more important is the fact that the shown Bell-type violations of local realism are robust.

Conceptually, these inequalities are related to earlier works on qubits \cite{allmeas,kz,nagata,lpn}. An exponentially strong violation of the inequalities will be demonstrated. 

We will utilize only measurements, for which no KS argument can be constructed. Note that therefore our results apply also to the case of locally contextual hidden variable theories.

{\em Observables}: Our construction will rely on squared spin-1 components $(\vec{S}\cdot\vec{n})^2$, where $\vec{n}$ is a unit vector. These observables are particularly interesting for their association to the sphere of Stern-Gerlach ``magnet" orientations. It should be stressed that this sphere is different from the Bloch sphere parameterizing qubit observables. In the case of squared components of spin-1 we have identification $\vec{n}\leftrightarrow-\vec{n}$. 

Since the square of the total angular momentum of a qutrit is 2, we have (we put $\hbar=1$)
\begin{equation}
S^2=(\vec{S}\cdot\vec{x})^2+(\vec{S}\cdot\vec{y})^2+(\vec{S}\cdot\vec{z})^2=2.
\end{equation}
 The three spin components mutually commute and thus obey {\bf the 1-0-1 rule}, which gives rise to all KS arguments for qutrits. That is, a simultaneous measurement of all three squared orthogonal spin-1 components will always give two 1's and one 0 in some order. 
 
 Note that one can introduce state $|\vec{n}\rangle$, for which $(\vec{S}\cdot\vec{n})^2|\vec{n}\rangle=0$. A kind of Malus law for spin-1 systems reads
\begin{equation}
|\langle \vec{n}|\vec{n}'\rangle |^2=(\vec{n}\cdot\vec{n}')^2.
\end{equation}

The whole set of squared spin-1 components is still too large for our purposes. There cannot exist hidden variables assigning outcomes to all measurements on the sphere. A simple proof of this statement was described in Ref. \cite{klyachko}. Bell's theorem would formally not even make a sense. Hence we focus only on observables parameterized by vectors $\vec{n}(\phi)=\frac{1}{\sqrt{3}}(\sqrt{2}\cos{\phi},\sqrt{2}\sin{\phi},1)$. In such a case $(\vec{S}\cdot\vec{n}(\phi))^2$ is represented by a matrix
\begin{equation}
S^2(\phi)=\frac{1}{3}\left(\begin{array}{ccc} 2&e^{-i\phi}&e^{-2i\phi}\\e^{i\phi}&2&-e^{-i\phi}\\e^{2i\phi}&-e^{i\phi}&2\end{array}\right).
\end{equation} 
Note that any three observables $\vec{S}\cdot\vec{n}(\phi_j)$ with $\phi_j=\phi_0+j\frac{2\Pi}{3}$, where $j=0,1,2$ constitute a triad of measurements at mutually orthogonal directions and hence their squares obey the 1-0-1 rule. Moreover, two different such trios, with $\phi_0-\phi_0'\neq j\frac{2\Pi}{3}$, do not share any eigenvectors and hence are completely independent from each other in a local realist's eyes.

{\em Construction:} 
We shall now demonstrate how the conflict between the local realistic and the quantum mechanical descriptions of entangled qutrits may arise.
The derivation is most concise if one uses instead of operators $S^2(\phi)$  traceless observables, $O(\phi)=S^2(\phi)-2/3$. Obviously, their eigenvalues are $1/3$ and $-2/3$.

 In every run of an experiment let each of observers 1,2,3,...,$N$ receive one spin-1 system from a common source. Each of them  measures a single squared spin component in direction $\vec{n}(\phi_1),\vec{n}(\phi_2),\vec{n}(\phi_3),...$, respectively (getting $\frac{1}3$ or $-\frac{2}3$, and thus obtains one of the eigenvalues of his/her observable $O(\phi)$. Subsequently they average the product of their results over many runs. This average will be called the (quantum mechanical) correlation function, $E_{QM}(\phi_1,\phi_2,\phi_3,...)$ and is an element of some real linear space of functions. In this space we introduce a standard scalar product,
\begin{equation}
(A,B)=\sum_{\begin{array}{c}\phi_1\in \Sigma_1\\\phi_2\in \Sigma_2\\...\end{array}}A(\phi_1,\phi_2,...)B(\phi_1,\phi_2,...)
\end{equation} 
in the discrete case ($\Sigma_1$, $\Sigma_2$,... are countable), or
\begin{eqnarray}
&(A,B)=\int_{\phi_1\in \Sigma_1}d\phi_1\int_{\phi_2\in \Sigma_2}d\phi_2...\nonumber\\
&\times A(\phi_1,\phi_2,...)B(\phi_1,\phi_2,...)
\end{eqnarray} 
in the continuous case ($\Sigma_1$, $\Sigma_2$,... are not countable). Hereafter, we will rather use the sum notation, but all considerations are valid for both cases. Here we focus only on the symmetric case of $\Sigma_1=\Sigma_2=...=\Sigma$. To take a benefit of the 1-0-1 rule, we assume that whenever some $\phi_0$ belongs to $\Sigma$, so do $\phi_0+\frac{2\pi}{3}$ and $\phi_0+\frac{4\pi}{3}$.

 The usual assumptions of local hidden variable theories lead to  following structure for correlation functions
\begin{equation}
E_{LR}(\phi_1,\phi_2,...)=\int d\lambda \rho(\lambda)\prod_{i=1}^{N}{I_i(\phi_i, \lambda)},
\end{equation}
where $\lambda$ represents hidden variables,  $\rho(\lambda)$ is their distribution, 
and each $I_i(\phi_i)$ takes values equal to the local eigenvalues, that is  $\frac{1}{3}$ or $-\frac{2}{3}$.
The ``unusual" additional assumption is that  $I_i(\phi_i)+I_i\left(\phi_i+\frac{2\Pi}{3}\right)+I_i\left(\phi_i+\frac{4\Pi}{3}\right)=0$, which expresses the 1-0-1 rule.
Such models form a convex set $\epsilon_{LR}$ (the additional assumption plays no role in the case of this property). For its extreme points all measurement outcomes are deterministic, and the whole correlation function $E_{LR}(\phi_1,\phi_2,...)$ factorizes into $\prod_{i=1}^{N}{I_i(\phi_i)}$. If the following inequality holds for all $E_{LR}\in\epsilon_{LR}$:  
\begin{equation}
\label{Bell1}
E_{QM}^2=(E_{QM},E_{QM})>(E_{QM},E_{LR}),
\end{equation}
then $E_{QM}\neq E_{LR}$, that is
a local realistic description is not possible. 
Notice that since the right-hand side of inequality (\ref{Bell1}) is linear in $E_{LR}$, its maximum will be attained for one of the extreme points.

{\em A simpler form of the inequality: }
One can simplify the expression for $\max_{E_{LR}}(E_{QM},E_{LR}).$ Let us consider the example of 2 spin-1 systems. Since $I_i(\phi_i)$ takes values $\frac{1}3$ and $-\frac{2}3$, 
we can write
$I_i(\phi_i)=\frac{1}{3} - \chi_i(\phi_i),$
where $ \chi_i(\phi_i)$ is a characteristic function giving values  $1$,  if $\phi_i$ belongs to the subset for which $I_i(\phi_i)=-\frac{2}{3}$, and otherwise  $0$.
Thus e.g.,
\begin{eqnarray}
&\max_{E_{LR}}(E_{QM},E_{LR})&\nonumber\\
&=\sum_{\phi_1\in\Sigma}\sum_{\phi_2\in\Sigma}E_{QM}(\phi_1,\phi_2)I_1(\phi_1)I_2(\phi_2)&\nonumber\\
&=\sum_{\phi_1\in\Sigma}\sum_{\phi_2\in\Sigma}E_{QM}(\phi_1,\phi_2)(\frac{1}{3} - \chi_1(\phi_1))(\frac{1}{3} - \chi_2(\phi_2))&\nonumber\\
\label{proof}
\end{eqnarray}
However,  $$\sum_{\phi_1\in\Sigma}E_{QM}(\phi_1,\phi_2)=\sum_{\phi_2\in\Sigma}E_{QM}(\phi_1,\phi_2)=0.$$ Simply from the 1-0-1 rule any trio of the form $E_{QM}(\phi_1,\phi_2),$ $E_{QM}\left(\phi_1+\frac{2}3\pi,\phi_2\right)$ and $ E_{QM}\left(\phi_1+\frac{4}3\pi,\phi_2\right)$ sums up to 0.  Thus the only term that does not  vanish is the one with the product of the characteristic functions. Therefore, we get
\begin{equation}
\label{lrmax}
\max_{E_{LR}}(E_{QM},E_{LR})=\sum_{\phi_1\in\sigma_1}\sum_{\phi_2\in\sigma_2}E_{QM}(\phi_1,\phi_2),
\end{equation}
where $\sigma_i$ is the set of values of $\phi_i$, for which $I_i(\phi_i)=-\frac{2}{3}$. All this generalizes to the multiparticle case in a trivial way. 

%The formula for $\max_{E_{LR}}(E_{QM},E_{LR})$ can be significantly simplified. Notice that the functions $I_i(\phi_i)$ take values $\frac{1}{3}$ and $-\frac{2}3$. Hence, each of them can be rewritten with a use of a characteristic function, $I_i(\phi_i)=\left(\frac{1}{3}-\chi_{i}(\phi_i)\right)$. $\chi_{i}(\phi_i)$ takes value $1$ if $I_i(\phi_i)=-\frac{2}{3}$ and vanishes otherwise. Take the example of two spins-1:
%\begin{eqnarray}
%&\max_{E_{LR}}(E_{QM},E_{LR})&\nonumber\\
%&=\sum_{\phi_1\in\Sigma}\sum_{\phi_2\in\Sigma}E_{QM}(\phi_1,\phi_2)I_1(\phi_1)I_2(\phi_2)&\nonumber\\
%&=\sum_{\phi_1\in\Sigma}\sum_{\phi_2\in\Sigma}E_{QM}(\phi_1,\phi_2)(\frac{1}{3} - \chi_1(\phi_1))(\frac{1}{3} - \chi_2(\phi_2)).&\nonumber\\
%\end{eqnarray}
%Now, we take the advantage from the 1-0-1 rule and the fact that $\Sigma$ contains only complete trios of measurements. These two facts 
%lead to a conclusion that $\sum_{\phi_1\in\Sigma}E_{QM}(\phi_1,\phi_2)=\sum_{\phi_2\in\Sigma}E_{QM}(\phi_1,\phi_2)=0.$ Thus only one term in the bottom line of Equation above has a nonzero contribution,
%\begin{eqnarray}
%&\max_{E_{LR}}(E_{QM},E_{LR})&\nonumber\\
%&=\sum_{\phi_1\in\Sigma}\sum_{\phi_2\in\Sigma}E_{QM}(\phi_1,\phi_2)\chi_1(\phi_1)\chi_2(\phi_2)&\nonumber\\
%&=\sum_{\phi_1\in\sigma_1}\sum_{\phi_2\in\sigma_2}E_{QM}(\phi_1,\phi_2),
%\end{eqnarray}
%where $\sigma_i$ is a subset of $\Sigma$, in which $I_i(\phi_i)=-\frac{2}3$. 

{\em Falsification of local realism:} We will demonstrate instances of violation of inequalities of the type (\ref{Bell1}). Let us focus on biased GHZ states,
\begin{equation}
\label{badany}
|\psi_N\rangle=\frac{2}{3}\left(|-1\rangle^{\otimes N}+\frac{1}2|0\rangle^{\otimes N}+(-1)^N|1\rangle^{\otimes N}\right),
\end{equation} 
where $|\pm 1\rangle$ and $|0\rangle$ are eigenstates of $S_z$ with respective eigenvalues $\pm 1$ and 0. The quantum mechanical correlation function for state (\ref{badany}) and observables $O(\phi)$ reads 
\begin{eqnarray}
&E_{QM}(\phi_1,\phi_2,...)\nonumber\\
&=\frac{8}{3^{2+N}}\left(\cos\sum_i\phi_i+(-1)^N\cos 2\sum_i\phi_i\right).
\end{eqnarray}

The values of $(E_{QM},E_{QM})$ for the continuous case, $\Sigma=[0,2\pi)$ read
\begin{equation}\label{CONT}
(E_{QM},E_{QM})=\frac{64(2\pi)^N}{3^{(4+N)}},
\end{equation}
whereas for a uniform distribution of $3n$ settings, $\Sigma=\left\{\phi:\phi=\frac{2j \pi}{3n},j=0,1,2,..3n-1\right\}$ we have 
\begin{equation}
\label{eqm}
(E_{QM},E_{QM})=\frac{64n^N}{3^{(4+N)}}
\end{equation}
(the  formula does not hold for the irrelevant case of $n=1$).

{\em Estimation of upper bound for local realistic models, for specific sets of settings:}
Here we are going to present the method to bound from above $(E_{QM},E_{LR})$. First, let us introduce
\begin{equation}
\phi_x^{kj}=\frac{2\pi}{3n}k+\frac{2\pi}{3}j
\end{equation}
as settings that we shall work with. Here $k=1,...,n$, $l=0,1,2$, whereas $x=1,2,...,N$ denotes the observer.
We want to estimate the maximal value of
\begin{eqnarray}
\label{theproblem}
&(E_{QM},E_{LR})\nonumber\\
&=(-1)^N\sum_{k_1=0}^{n-1}\sum_{j_1=0}^2...\sum_{k_N=0}^{n-1}\sum_{j_N=0}^2\nonumber\\
&E(\phi_1^{k_1j_1},...,\phi_N^{k_Nj_N})\chi_{\sigma_1}(\phi_1^{k_1j_1})\chi_{\sigma_2}(\phi_2^{k_2j_2})...\chi_{\sigma_N}(\phi_N^{k_Nj_N})\nonumber\\
\end{eqnarray}
$\chi_{\sigma_x}(\phi_x)$ is the characteristic function, equal to 1 when $\phi\in\sigma_i$ and 0 otherwise. Note that for any $k$ there is always only one $j$ for which $\chi_{\sigma_x}(\phi_x)=1$. This is a manifestation of the 1-0-1 rule.

We can decompose the correlation function according to well-known expressions for trigonometric functions.
\begin{eqnarray}
&E_{QM}(\phi_1,...,\phi_N)\nonumber\\
&=\frac{8}{3^{2+N}}(\cos\phi_1\cos\phi_2\cos\phi_3...\cos\phi_N...\nonumber\\
&+(-1)^N(\cos 2\phi_1\cos 2\phi_2...\cos 2\phi_N+...)),
\end{eqnarray}
where 
\begin{equation}
\cos\psi_1\cos\psi_2\cos\psi_3...\cos\psi_N+... \label{SEQUENCE}
\end{equation}
stands for the sequence forming the decomposition of  $\cos\sum_i\psi_i$. As we shall see, the actual details of such a decomposition will be irrelevant. What is important is that we have a certain sequence of terms in form of product of trigonometric functions, each one containing either $cos\psi_k$ or $sin\psi_k$. Each term  is multiplied by a coefficient $\pm1$. 

Notice that the values of  $\cos\phi_x^{kj}$, $\sin\phi_x^{kj}$ $\cos 2\phi_x^{kj}$, and  $\sin 2\phi_x^{kj}$ form four orthogonal $3n$-dimensional vectors $\vec{c}_{1}$, $\vec{s}_{1}$, $\vec{c}_{2}$, and $\vec{s}_{2}$, respectively, which all have the same equal norm, denoted here as $M$. Hence we can put problem (\ref{theproblem}) into the following form
\begin{eqnarray}
\label{vectorform}
&\max_{E_{LR}}(E_{QM},E_{LR})\nonumber\\
&=\frac{8}{3^{2+N}}\max_{\sigma_1,..\sigma_N}(-1)^N[\vec{c}_{1}\otimes\vec{c}_{1}\otimes...\otimes\vec{c}_{1}+...\nonumber\\
&+(-1)^N(\vec{c}_{2}\otimes\vec{c}_{2}\otimes...\otimes\vec{c}_{2}+...)]\cdot\vec{\chi}_1\otimes...\otimes\vec{\chi}_N,\nonumber
\end{eqnarray} 
where $\vec{\chi}_x$ is a $3n$ dimensional vector with entries $0,1$ corresponding to the characteristic function $\chi_{\sigma_x}(\phi^{kj}_x)$, and $\cdot$ represents the standard real ``dot" scalar product in a $3nN$ dimensional real space. The compound symbol  
$\vec{c}_l\otimes\vec{c}_l\otimes...\otimes\vec{c}_l+...$ represents a sequence of tensor products in obvious relation with the former sequence of trigonometric functions (\ref{SEQUENCE}). That is every $\cos$ is replaced by $\vec{c}_l$, every $\sin$ by $\vec{s}_l$, and the coefficients of the expansion are kept unchanged.

Now  we use an important technical observations. Fist we notice that, 
\begin{eqnarray}
&(\vec{c}_{l}\otimes\vec{c}_{l}\otimes...\otimes\vec{c}_{l}+...)\cdot\vec{\chi}_1\otimes...\otimes\vec{\chi}_N&\nonumber \\
&= (\vec{c}_{l}\cdot \vec{\chi}_1)( \vec{c}_{l}\cdot \vec{\chi}_2)...(\vec{c}_{l}\cdot\vec{\chi}_N)+...,&\nonumber \\
\end{eqnarray}
where now $\cdot$ stands for a scalar product in a real $3n$ dimensional space.
Since the four vectors, $\vec{c}_{1}$, $\vec{s}_{1}$, $\vec{c}_{2}$, and $\vec{s}_{2}$, are orthogonal, we can always put
\begin{eqnarray}
&\vec{\chi}_{x}\cdot\vec{c_{l}}=||\vec{\chi}_x^{||}||M\cos\beta_x^l\cos\left(\alpha_x+l\frac{\pi}{2}\right),\\
&\vec{\chi}_{x}\cdot\vec{s_{(l)}}=||\vec{\chi}_x^{||}||M\sin\beta_x^l\cos\left(\alpha_x+l\frac{\pi}{2}\right),
\end{eqnarray}
where $l=1,2$ and $\alpha_x$ and $\beta_x^S$ are some angles, and $||\vec{\chi}_x^{||}||$ denotes the norm of a projection of $\vec{\chi}_x$ onto the 4-dimensional space spanned by $\vec{c}_{1}$, $\vec{s}_{1}$, $\vec{c}_{2}$, and $\vec{s}_{2}$. This is an obvious property of any decomposition of a projection onto a 4 dimensional subspace. If whatever vector $\vec{\chi}$ is projected into a subspace  spanned by {\em normalized} basis vectors $\vec{a}_1$, $\vec{b}_1$, $\vec{a}_2$, $\vec{b}_2$, one can always find such three angles $\beta_1$, $\beta_2$ and $\alpha$ so that one can put
\begin{equation}
\vec{\chi}^{||}=||\vec{\chi}^{||}|| \sum_{l=1}^2\cos(\alpha + \frac{\pi}{2}l)(\cos\beta_l\vec{a}_l+\sin\beta_l\vec{b}_l).    
\end{equation}

After putting the above formulas to (\ref{vectorform}) we get the following
\begin{eqnarray}
\label{cosi}
&(E_{QM},E_{LR})\leq\frac{8}{3^{2+N}} (M||\vec{\chi}^{||}_x||)^N\nonumber\\
&\times\max \big[\prod_{i=1}^N\cos\alpha_i\cos(\sum_{j=1}^N\beta^1_j)\nonumber\\
&+(-1)^N\prod_{i=1}^N\cos(\alpha_i+\frac{\pi}{2})\cos(\sum_{j=1}^N\beta^2_j)\big]\nonumber\\
\end{eqnarray}
The  functions of $\sum_{i=1}^N\beta_i^l$ result from a reverse application of the expansion formula for trigonometric functions of many angles: $\cos\beta_1^l\cos\beta_2^l\cos\beta_3^l...\cos\beta_N^l+...=\cos(\sum_{i=1}^N\beta_i^l) $.
Since cosine  is never greater than $1$, the expression in the square bracket can be majorized by $\left|\prod_{i=1}^N\cos\left(\alpha_i\right)\right|+\left|\prod_{i=1}^N\sin\left(\alpha_i\right)\right|$, which in turn can be seen as a scalar product of two vectors, $(\pm \cos\alpha_1,\pm \sin\alpha_1)$ and $\left(\prod_{i=2}^N\cos\left(\alpha_i\right),\prod_{i=2}^N\sin\left(\alpha_i\right)\right)$, both with norm not exceeding 1. By the Cauchy inequality, the value of this expression cannot exceed 1. therefore we get 
\begin{eqnarray}
(E_{QM},E_{LR})
\leq\frac{8}{3^{2+N}}(M||\vec{\chi}^{||}_x||)^N \label{BOUND}
\end{eqnarray}
Thus once we know for the maximal possible value of $||\vec{\chi}^{||}_x||M$, we can estimate the local realistic bound from above (note that the above bound does not have to be tight).

 For $n=3$ we found it by a MATHEMATICA case-by-case study to be
\begin{eqnarray}
&\max_{\sigma_i}||\vec{\chi}_i^{||}||M=\sqrt{\left(1+2\cos\frac{2\pi}{9}\right)^2+\left(1+2\sin\frac{\pi}{18}\right)^2}\nonumber\\
&\approx 2.86822.
\end{eqnarray}
Hence this particular estimate can be used for an arbitrary even $N$. The strength of violation can be then estimated from Eq. (\ref{eqm}) as
\begin{eqnarray}
\label{bound}
&\frac{(E_{QM},E_{QM})}{\max_{E_{LR}}(E_{QM},E_{LR})}\nonumber\\
&\leq\frac{8}{9}\left(\frac{3}{\sqrt{\left(1+2\cos\frac{2\pi}{9}\right)^2+\left(1+2\sin\frac{\pi}{18}\right)^2}}\right)^N,
\end{eqnarray}
which predicts violation of the inequality for $N\geq 3$. We have found the values of $\max_{\sigma_i}||\vec{\chi}_i^{||}||M$ for $n=4,5,6,7$ to be $3.7678$, $4.678$, $5.5932$, and $6.5112$, respectively. Note that these values are smaller than $n$, which when one compares   (\ref{BOUND}) and (\ref{eqm}) guaranties falsification of Local Realism for sufficiently large $N$. Interestingly, for $N=2$ the bound given by Ineq. (\ref{BOUND}) is tight for all $n$.  

{\em Numerical analysis:}

For $N=2$ we have numerically tested all local realistic models for $n<7$ and verified that the highest values of $(E_{QM},E_{LR})$ for $n=2,3,4,5,6$ are 0.1975, 0.6083, 1.1852, 1.6632, and 2.380, which corresponds to violation ratios $\frac{1}{V}=\frac{(E_{QM},E_{QM})}{\max_{E_{LR}}(E_{QM},E_{LR})}$ equal to 0.8889, .9724, 1.0018, 1.016, and 1.023, respectively. $V$ tells us that to what extend we can  maintain the violation, while we mix the state $|\psi\rangle \langle \psi|$ with the white noise, $\rho_{noise}$ (proportional to the unit operator), so that we get a mixed state $P|\psi\rangle \langle \psi|+(1-P))\rho_{noise}$. In such a case $P=V$ is the threshold value at which the inequalities are not violated anymore. Please note that the correlation functions for the noise state vanish. 

\begin{figure}[t]
\includegraphics[width=6cm]{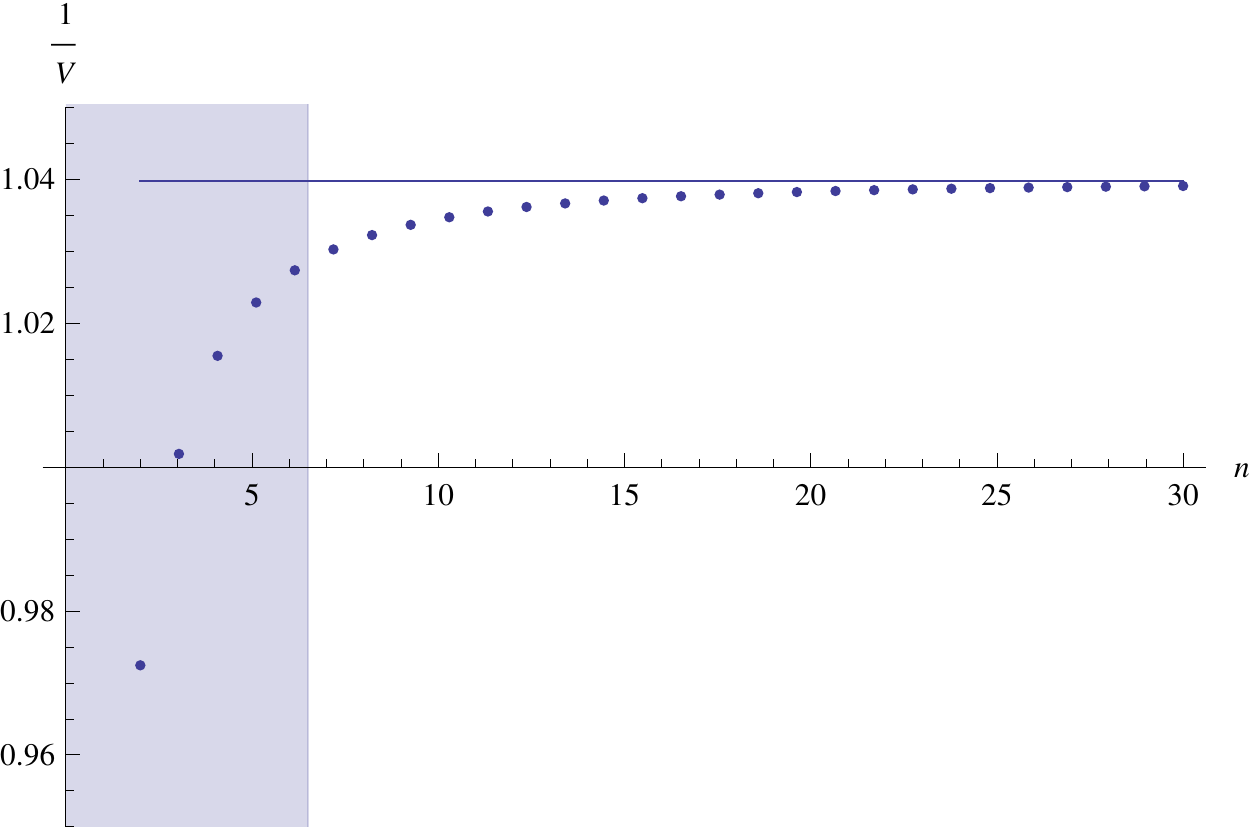}
\caption{The violation ratio $\frac{1}{V}$ in function of the number of local measurement trios $n$ for 2 spin-1 systems. In {\em all figures} the solid line shows the continuous limit. In the shaded region the results are calculated explicitly, outside this region we show an estimate based on the conjecture (see text for details).}
\end{figure}

\begin{figure}[t]
\includegraphics[width=6cm]{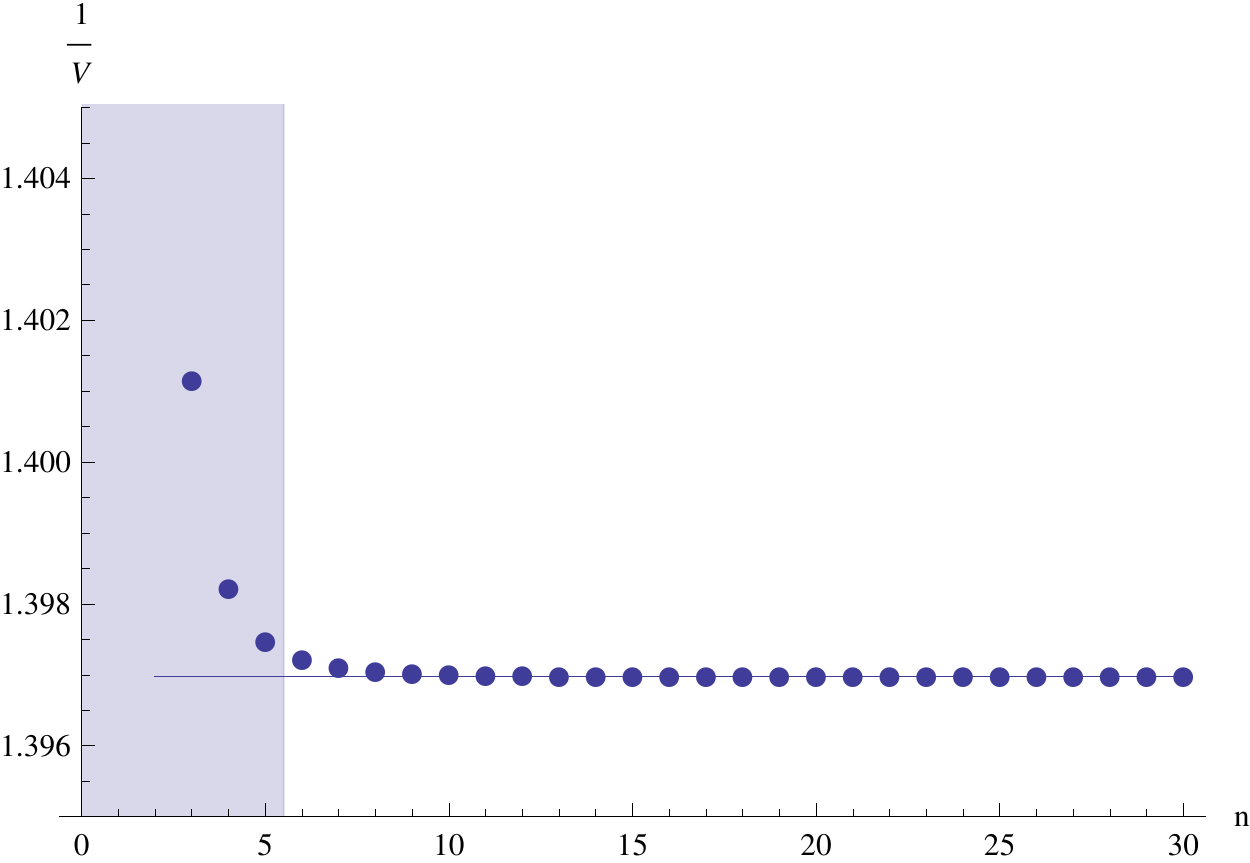}
\caption{The violation ratio $\frac{1}{V}$ in function of the number of local measurement trios $n$ for 3 spin-1 systems. }
\end{figure}

\begin{figure}[t]
\includegraphics[width=6cm]{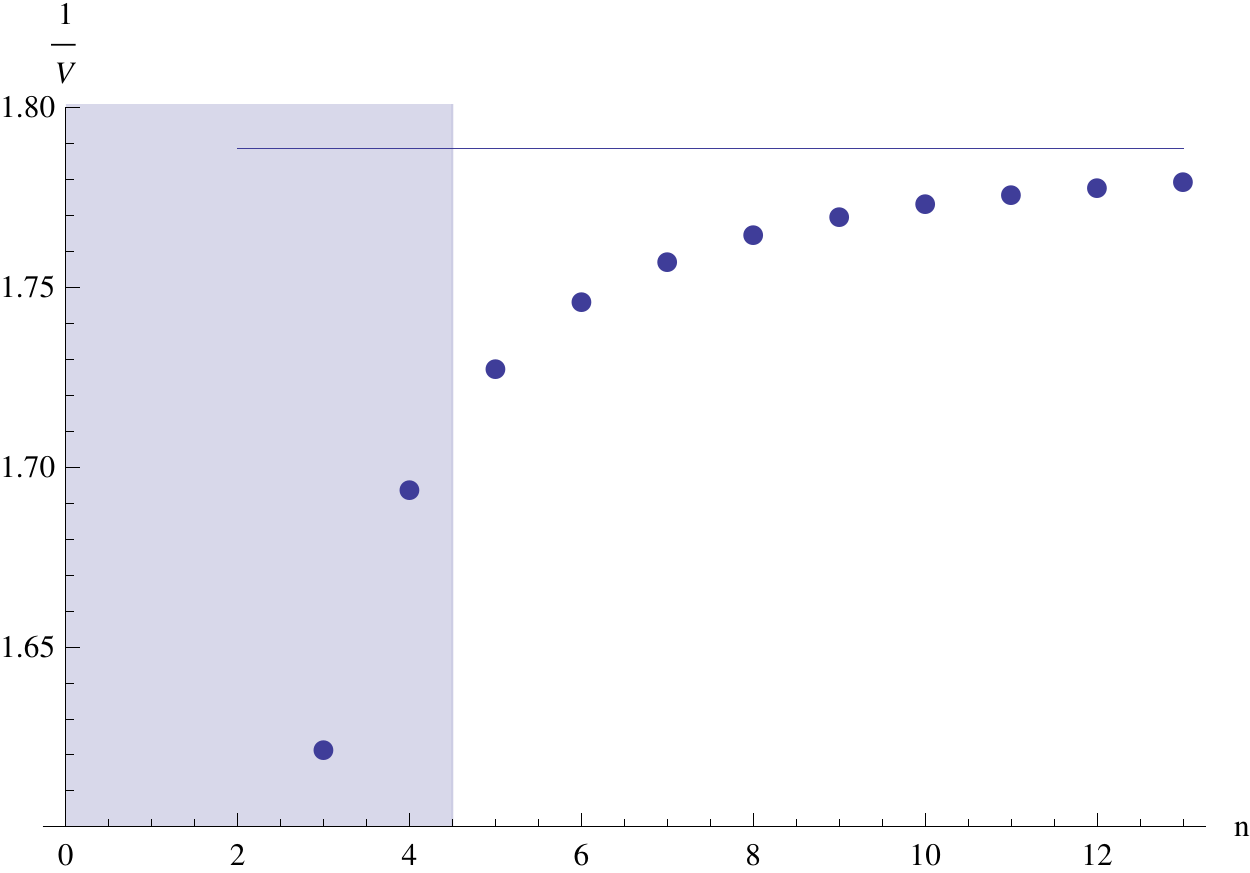}
\caption{The violation ratio $\frac{1}{V}$ in function of the number of local measurement trios $n$ for 4 spin-1 systems. }
\end{figure}

\begin{figure}[t]
\includegraphics[width=6cm]{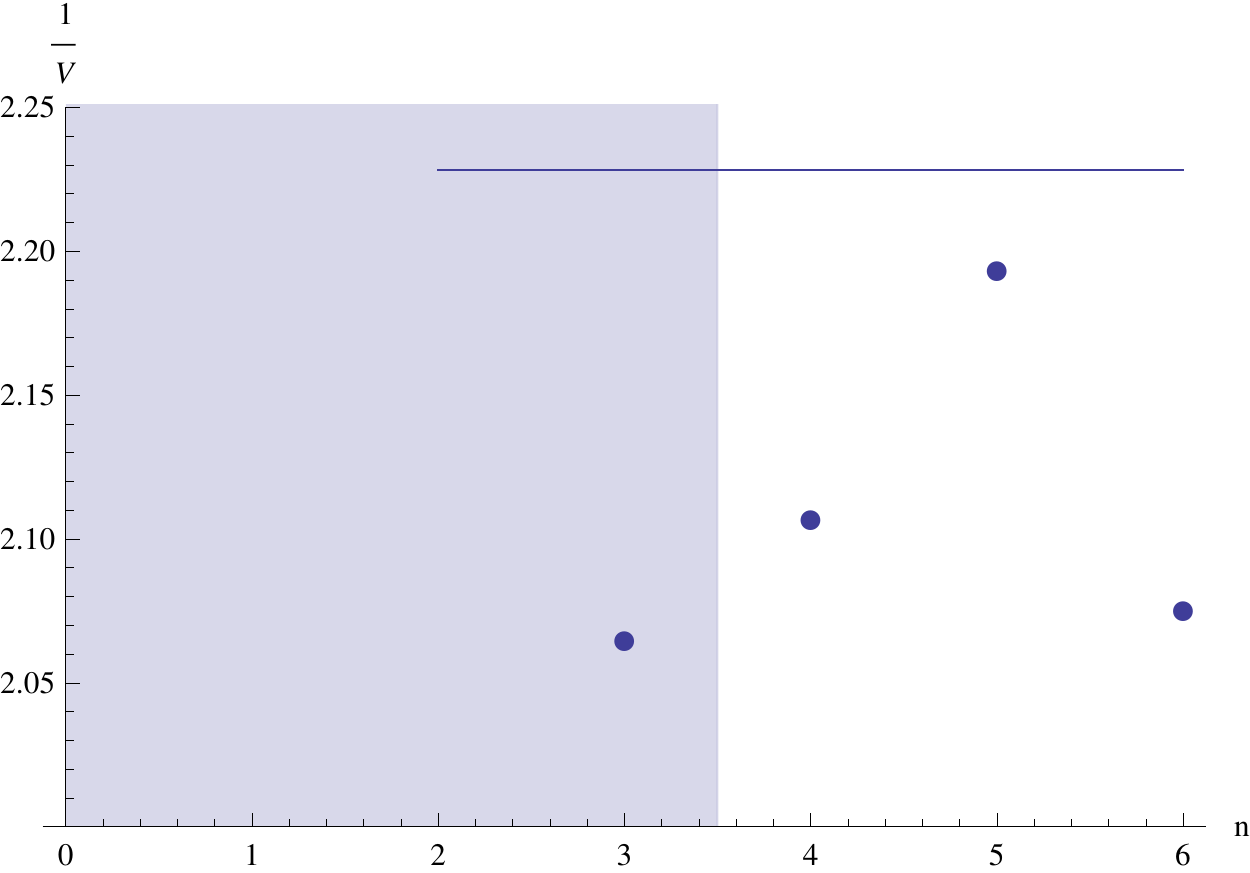}
\caption{The violation ratio $\frac{1}{V}$ in function of the number of local measurement trios $n$ for 5 spins-1 systems. }
\end{figure}

Similarly we processed {\em all} local realistic models for $N=3$ and $n=2,3,4,5$. We have found that $\frac{1}{V}$ equals 0.889, 1.401, 1,398 and 1.397 respectively.% where the first value (which does not falsify the local realistic description) is attained for $\sigma$'s of type $\sigma''$, whereas all others appear for $\sigma_j$'s of the first type. Again, we take this as a conjecture and estimate the violation for $n$ up to 30. 
The results are presented in Fig 2.

{\em Conjecture}: 
All these numerically analyzed instances reveal a specific pattern of the form of the optimal sets $\sigma$ (that is those that lead to the highest value of (\ref{lrmax})). 
There are always many local realistic models maximizing the scalar product with the quantum mechanical function, but by the symmetry of $E_{QM}$ (which is solely a function of $\sum_{i=1}^N\phi_i$) one of them has $\sigma_1=\sigma_2=...=\sigma_{N-1}=\sigma$ and $\sigma_N=R_a(\sigma)$. $R_a$ is an operator, which shifts its argument by $a$ modulo $2\pi$, $R_a(b)=a+b$. Two kinds of $\sigma$s are particularly relevant. One, denoted as $\sigma'$, has all points bunched together, e.g. $\sigma=\left[0,\frac{4\pi}{3}\right)$, or $\sigma=\left\{0,\frac{2\pi}{9},...,\frac{10\pi}{9}\right\}$. The other, called $\sigma''$, has two equally sized parts with possibly equally sized gaps, e.g., $\sigma=\left[0\,\frac{2\pi}{3}\right)\cup \left[\pi,\frac{5\pi}{3}\right)$ or  $\sigma=\left\{0,\frac{2\pi}{9},\frac{4\pi}{9},\frac{10\pi}{9},\frac{12\pi}{9},\frac{14\pi}{9}\right\}$. 

In all explicitly studied cases the optimal sets $\sigma_j$ for the local realistic value were of types either $\sigma$ or $\sigma''$. This observation can be taken as a conjecture for more complicated situations. For $N=2,3$ we have been able to give estimates of maximal violation under this assumption to up to 30 trios of observables. This can be seen in Figs. 1 and 2 and as points outside the shaded region

For higher $N$ we mainly used $\sigma_i$'s  of  either $\sigma'$ or $\sigma''$ type. The results are presented in Figs. 3 and 4. We have been able to estimate violations for $N\leq 6$. This hypothesis was additionally supported by studying {\em all} local realistic models for $N=4, n=2,3,4$ and $N=5, n=2,3$. %The general trend is that, as we predicted, sets of type $\sigma'$ maximizes the overlap between local realistic and quantum mechanical correlation function.

Notice that the points in the plot for $N=5$ are distributed in a different way  than in other plots. This is because three of four of these points are associated with $\sigma_i$'s of the second type. Calculations show that the plot of the the ratios of $(E_{QM},E_{QM})$ to maximal local realistic values under the restrictions that $\sigma_i$'s are of these type has some kinks for $mod(n,4)=1$. In Fig. 4 the third point corresponds to first such kink, but this value is overtaken with a value for $\sigma_i$'s of the first type. 

These observations we can be extended to the continuous set of observables. One can take $\sigma_i=\left(-\frac{2\pi}3,\frac{2\pi}3\right]$ ($i\neq N$) and $\sigma_N=\left(-\frac{2\pi}3+a,\frac{2\pi}3+a\right]$. 
\begin{eqnarray}
&(E_{QM},E_{LR})=\int_{-\frac{2\pi}{3}}^{\frac{2\pi}{3}}d\phi_1...\int_{-\frac{2\pi}{3}}^{\frac{2\pi}{3}}d\phi_{N-1}\int_{-\frac{2\pi}{3}+a}^{\frac{2\pi}{3}+a}d\phi_{N}&\nonumber\\
&\times\frac{8}{3^{N+2}}\left(\cos\sum_j\phi_j-\cos2\sum_j\phi_j\right)&\nonumber\\
&=\frac{\cos \left(a-\left(\frac{N}{3}-1\right)\pi\right)}{3^{\frac{N}2+2}2^{N-3}}
+\frac{\cos 2\left(a-\left(\frac{N}{3}-1\right)\pi\right)}{3^{\frac{N}2+2}8},&\nonumber\\
\end{eqnarray}
while for even $N$ we have
\begin{eqnarray}
&(E_{QM},E_{LR})=\int_{-\frac{2\pi}{3}}^{\frac{2\pi}{3}}d\phi_1...\int_{-\frac{2\pi}{3}}^{\frac{2\pi}{3}}d\phi_{N-1}\int_{-\frac{2\pi}{3}+a}^{\frac{2\pi}{3}+a}d\phi_{N}&\nonumber\\
&\times\frac{8}{3^{N+2}}\left(\cos\sum_j\phi_j+\cos2\sum_j\phi_j\right)&\nonumber\\
&=\frac{\cos\left(a-\frac{2N}{3}\pi\right)}{3^{\frac{N}2+2}8}+
\frac{\cos 2\left(a-\frac{2N}{3}\pi\right)}{3^{\frac{N}2+2}2^{N-3}}.\nonumber&\\
\end{eqnarray}
The maximum of these functions is $2^{3-N}3^{-\frac{N}2-2}(2^N+1)$. This, when compared with (\ref{CONT}) corresponds to a violation ratio $\frac{1}{V}=\frac{8}{9(2^N+1)}\left(\frac{4\pi}{3\sqrt{3}}\right)^N$. However, this is only a conjecture.

{\em Closing remarks:} We have proposed a new family of Bell-type inequalities for arbitrarily many spin-1 systems. Observers might have a divisible by three, but otherwise an arbitrary number of observables per side. These inequalities totally avoid the Kochen-Specker contradiction, but still use the $1-0-1$ rule, which is the source of it. Despite specific constraints that we introduced to our analysis,  violations of these inequalities are quite high, and grow exponentially with the number of spin systems.

On the trail of the derivation of the inequality we have tried to introduce some free parameters. For example, we have given up the tracelessness of the observables. However this only decreased the strength of violation. Similarly, changing the the degree of the biasedness of the state (i. e., the relative weights of its three components) had only a negative effect on the strength of violation. 

We thank \v{C}aslav Brukner for stimulating discussions. AD is supported within the International PhD Project "Physics of future quantum-based information technologies": grant MPD/2009-3/4 of Foundation for Polish Science. MW was supported by the program ``Optimization of Quantum Resources'' of the Ministry of Science and Higher Education of Poland and later the progam QUASAR of the National Centre for Research and Development of Poland. MZ is supported by QESSENCE project (VII FP EU).

\end{document}